# Theoretical evidence for new adsorption sites of $CO_2$ on the Ag electrode surface

Shuai Guo, [a,b] Yao Li, [a] Lei Liu, [a *] Xiangping Zhang, [a,c] Suojiang Zhang, [a *]

[a] CAS Key Laboratory of Green Process and Engineering, State Key Laboratory of Multiphase Complex Systems, Beijing Key Laboratory of Ionic Liquids Clean Process, Institute of Process Engineering, Chinese Academy of Sciences, Beijing 100190, China

[b] School of Future Technology, University of Chinese Academy of Sciences, Beijing 100049, China

[c] Dalian National Laboratory for Clean Energy, Dalian 116023, China

**ABSTRACT**: Nowadays, electrochemical reduction of $CO_2$ has been considered as an effective method to solve the problem of global warming. The primary challenge in studying the mechanism is to determine the adsorption states of $CO_2$, since complicated metal surfaces often result in many different adsorption sites. Based on the density functional theory (DFT) calculations, we performed a theoretical study on the adsorption of $CO_2$ on the Ag electrode surface. The results show that the adsorption populations of $CO_2$ are extremely sensitive to the adsorption sites. Importantly, we found that the preferable adsorption positions are the terrace sites, rather than the previous reported step sites. The adsorption populations were found with the order of (211) > (110) > (111) > (100). Subsequently, the adsorption characteristics were correlated with the *d*-band theory and the charge transfers between Ag surfaces and $CO_2$.

**TOC**: Density functional theory calculations realized that the most stable adsorption positions of $CO_2$ on Ag surfaces are the terrace sites, rather than the previous reported step sites, and (211) facet is the most populated surface.

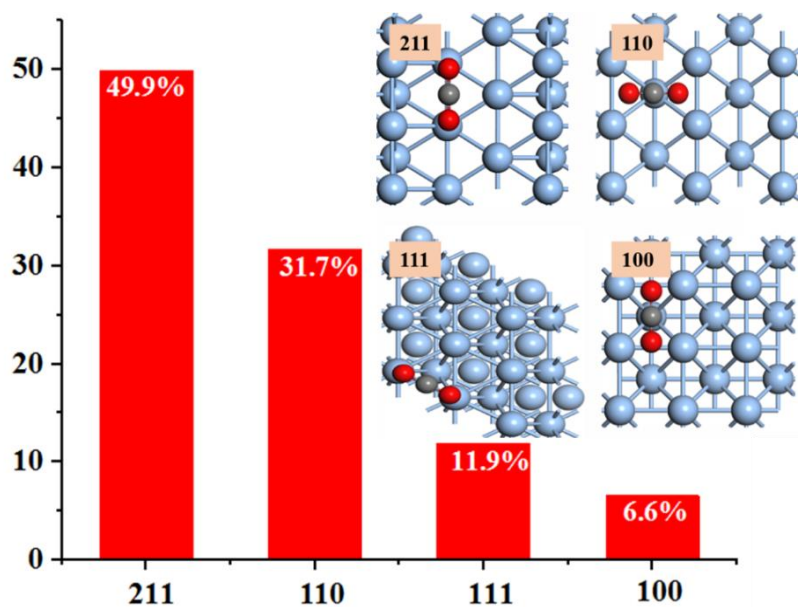

INTRODUCTION

Converting the $CO_2$ into valuable fossil fuels not only resolve the energy shortage, but also alleviates the global warming.[1] Researchers have come up with a variety of methods and techniques to convert $CO_2$.[2-4] Examples include but not limited to electrochemical, photocatalytic, thermochemical and enzymatic reduction of $CO_2$. Among them, the electrochemical reduction is one of the most promising methods, which has attracted extensive interests from researchers because of its high product selectivity and mild reaction conditions.[5-7] With this technique, $CO_2$ can be reduced into important products, such as CO, formic acid (HCOOH), ethanol ($CH_3CH_2OH$) and ethylene ($CH_2CH_2$), etc.[8-14] For example, Rosen et al.[11] reported the Faraday efficiency (FE) of 96% for $CO_2$ conversion to CO in the acidic solution (sulfuric acid and 1-ethyl-3-methylimidazoliumtetra-fluoroborate, [Emim][$BF_4$]) with a silver (Ag) electrode. Hollingsworth et al.[15] reported that the main product was HCOOH (FE>93%) with the same Ag electrode but using a superbase solvent (tetraalkyl phosphonium, [$P_{66614}$][124Triz]). Ma et al.[16] found that the sulfur-doped indium catalyst exhibited a high FE of HCOOH (>85%), a wide current density range (25~100 mA $cm^{-2}$), and a high generation rate of the HCOOH (1449 mol $h^{-1}$ $cm^{-2}$). Interestingly, Sargent et al. found that by modifying the catalyst surface with organic molecules (aromatic substituted bipyridine), the FE of converting $CO_2$ to $CH_3CH_2OH$ reach 41%,[13] which was further enhanced to 43 % if a catalyst of ($Ce(OH)_x$-doped-Cu) was employed.[6] With the same methodology, Sargent et al.[14] obtained a FE of $CO_2$ converting to $CH_2CH_2$ being 72% (current density 230 mA/$cm^2$).

Meanwhile, the mechanism of electrocatalytic $CO_2$ reduction reaction ($CO_2$RR) has been extensively investigated.[17-20] On this aspect, understanding $CO_2$ adsorption behaviors on electrode surfaces is an essential step, which attracts a large number of studies.[21-25] According to the geometric characteristics and electronic structures of $CO_2$, its adsorption state generally can be classified into chemical and physical adsorption[26-28], which is strongly depend on the types of metal electrodes.[29-31] Wang et al. found that $CO_2$ is physically adsorbed on the Au (111) and Ag (111) surfaces, while it has chemical interactions with Fe, Co, Ni, Cu, Rh, Pd and Pt electrodes (*i.e.* binding energies range from -0.68 to 1.01 eV, and the charge transfer between these metals and the $CO_2$ is between -0.37 and -0.83 *e*).[28] The adsorption of $CO_2$ on different surfaces of the same electrode can be also different.[27, 32] For example, Solymosi et al.[29] found that $CO_2$ decomposes on Fe (111) and Fe (100) surfaces. However, there is no decomposition of $CO_2$ on the Fe (110) surface based on the ultraviolet photoelectron spectroscopy.

For the conversion of $CO_2$ to CO, Ag is the most often used electrode.[11, 33-35] Although there have been many experimental [36-38] and theoretical studies [28, 39] on the adsorption of $CO_2$, the preferable adsorption site is still under debate. For example, Zhang et al.[39] reported the adsorption energies of $CO_2$ are -0.01, -0.01, -0.02, and -0.02 eV on (100), (111), (110), and (211) surfaces, respectively. However, Ko et al.[20] reported the adsorption energies of $CO_2$ are -0.26 eV at *bridge*, *hcp* and *fcc* sites on the (111) surface, and the adsorption energies of $CO_2$ at the equivalent sites on the (111) surface reported by Dietz et al.[40] were also -0.26 eV. Moreover, similar to other metals (*i.e.* Ru, Pt, Mo), Ag also has multiple adsorption sites on different surfaces.[40-42] However, it is still not clear that which is the most stable adsorption site. For this vein, we performed DFT calculations with the goals as follows: 1) to find the most stable adsorption site, and 2) to identify the most populated surface of $CO_2$. Through this study, we obtained new adsorption sites of $CO_2$, and the populations of $CO_2$ on different Ag surfaces. Subsequently, the *d*-band theory, the electronic structures of $CO_2$ and Ag surfaces were employed to understand the stabilities of the surfaces and the adsorption behaviors.

COMPUTATIONAL METHODS

The surfaces of (211), (110), (111) and (100) were selected to study the adsorption of $CO_2$. The experimental lattice constants (4.085 Å) was adopted, which is similar to the previous computational studies.[43-45] Our models contained a 3 × 3 supercell with four Ag layers, of which the bottom two layers were fixed during the optimizations. A vacuum of 15 Å was added above Ag surfaces. The density functional theory calculations were performed by employing the *Vienna ab initio* simulation package (VASP).[46] The generalized gradient approximation (GGA) Perdew–Burke–Ernzerhof (PBE) functional was used to describe the electronic structure properties of molecules and surfaces.[47] For ion relaxation, the force convergence was less than 0.05 eV/Å, and the energy convergence was $10^{-5}$ eV. The plane-wave basis cutoff was set to 450 eV. The D3 correction of Grimme was added to all calculations.[48] The $\Gamma$-centered $k$-point Monkhorst−Pack grid was set to be 1 × 1 × 1 for relaxation, while 12×12×1 $k$-point was applied for $d$-band center calculations. The Bader charge was calculated by the code of the Henkelman group.[49] Factors that may affect the adsorption energies were benchmarked, including system sizes, lattice parameters, $k$-points, and energy cut-off, which are provided in the Supporting Information.

The adsorption energies of $CO_2$ ($E_{ads}$) are calculated by the equation 1:

$$E_{ads} = E(CO_2/slab) - [E(CO_2) + E(slab)] \qquad (1)$$

where $E(CO_2/slab)$ is the total electronic energy of the Ag surfaces and $CO_2$, $E(CO_2)$ is the electronic energy of $CO_2$, and $E(slab)$ is the electronic energy of the clean Ag surface.

According to Boltzmann distributions, the population of $CO_2$ at each adsorption site are calculated *via* equation 2.[50]

$$P_i = \frac{e^{-\varepsilon_i/kT}}{\Sigma_{j=1}^{M} e^{-\varepsilon_j/kT}} \tag{2}$$

where $P_i$ is the probability of particle distribution at position $i$, $\varepsilon_i$ is the adsorption energy at position $i$, $k$ is the Boltzmann constant, and $T$ is the temperature (298.15 K).

RESULTS AND DISCUSSION

The adsorption sites of $CO_2$ on Ag surfaces are listed in Fig. 1(a) ~ (d). Specifically, fourteen adsorption sites are found on the (211) surface, including three *top* sites (T1, T2, and T3), six *bridge* sites (B1, B2, B3, B4, B5, and B6), and five *hollow* sites (H1, H2, H3, H4, and H5). The (110) surface has seven adsorption sites, including two *top* sites (T1 and T2), three *bridge* sites (B1, B2, and B3) and two *hollow* sites (H1 and H2). The (111) surface has four adsorption sites, which are one *top* (T), one *bridge* (B), one *two-fold hollow* (H) and one *three-fold hollow* (F), respectively. There are three adsorption sites on the surface of (100), which are one *top* (T), one *bridge* (B), and one *two-fold hollow* (H). On the other hand, $CO_2$ has four, four, seven and three possible orientations on (211), (110), (111) and (100) surfaces, respectively (see Fig. 2, and Fig. S1~ S3). In total, we found 56, 28, 13 and 9 structures of $CO_2$ adsorbed on (211), (110), (111) and (100) surfaces, respectively. For clarify, we employed labels for each structure with a rule as follows: preceding letters (including numbers) represent adsorption sites of $CO_2$, and the last letter (*a*, *b*, *c*, and *d*) represent the orientation of $CO_2$. For example, T1a refers to the T1 site (Fig.1) with *a* orientation of $CO_2$ (Fig.2).

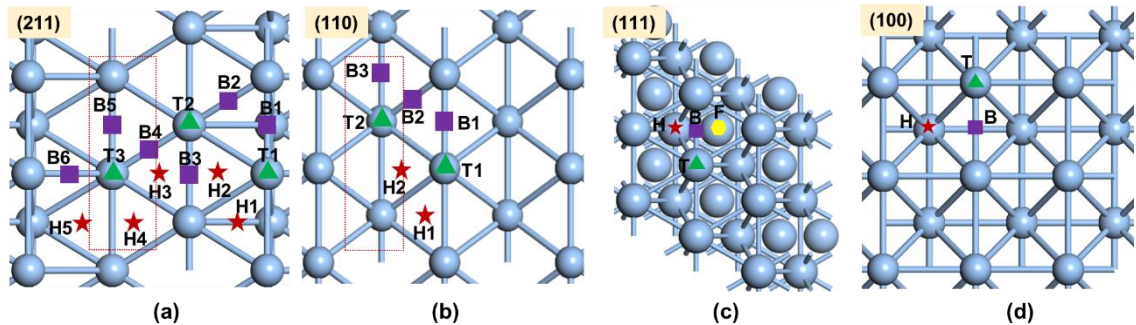

**Figure. 1**. Top views of (a) (211), (b) (110), (c) (111), (d) (100) surfaces. The red dashed line in (a) and (b) are for the convenience of later discussion.

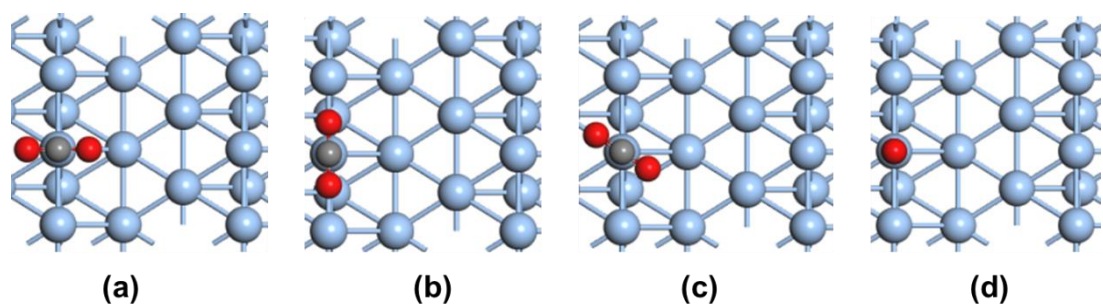

**Figure. 2**. Top views of $CO_2$ orientations at the T1 site of the (211) surface, and top view of $CO_2$ orientations on other surfaces are shown in Fig. S1~S3. Color legend: Ag grey, C black and O red.

According to previous studies, [39] initial distances between $CO_2$ and the (211) surface were set as 3.3 Å, and $CO_2$ were set to be linear ($\angle OCO=180°$). The selected optimized structures and corresponding adsorption energies are shown in Fig. 3 (for all structures and their adsorption energies, see Fig. S4 and Table S1). Of these structures, B5b and H4c have the largest adsorption energies (*i.e.*, -0.29 eV). The distances between $CO_2$ and nearest Ag atom are $d_{C-Ag}$ = 3.7 Å, $d_{O1-Ag}$ = 3.6 Å, $d_{O2-Ag}$ = 3.6 Å for B5b, and $d_{C-Ag}$ = 3.6 Å, $d_{O1-Ag}$ = 3.5 Å, $d_{O2-Ag}$ = 3.6 Å for H4c, respectively. The next two meta-stable structures are T3a and B4c, and their adsorption energies are -0.28 eV. The distances from O and C to the nearest Ag are $d_{C-Ag}$ = 3.9 Å, $d_{O1-Ag}$ = 3.6 Å, $d_{O2-Ag}$ = 3.9 Å for T3a, and $d_{C-Ag}$ = 3.6 Å, $d_{O1-Ag}$ = 3.4 Å, $d_{O2-Ag}$ = 3.7 Å for B4c, respectively. Note that Zhang et al. [39] reported the adsorption energy of $CO_2$ is -0.02 eV when Van der Waals (vdW) corrections are not included in the DFT calculations. We also computed the adsorption energies of $CO_2$ at T1, T2, T3 and B1 without vdW corrections, which are -0.02, -0.03, -0.02 and -0.02 eV, respectively (Table S3). Besides the geometric characteristics of the $CO_2$, Bader charge analysis also confirm that $CO_2$ is physically adsorbed on the (211) surface. For example, the charge transfers between

CO$_2$ and the (211) surface are only 0.061 $e$ for B5b, 0.050 $e$ for T2a, and 0.041 $e$ for T1b, respectively. Notably, we found that the adsorption energies of CO$_2$ vary largely with respected to the adsorption sites. Generally, the terrace sites (B5, B4, H4 and T3, highlighted with the red dot box in Fig.1a) have larger values than that of the step (*i.e.* B1, B6, H1 and T1). For example, the adsorption energy of B5b is -0.29 eV, while the adsorption energy of T1b is -0.15 eV (Fig. 3).

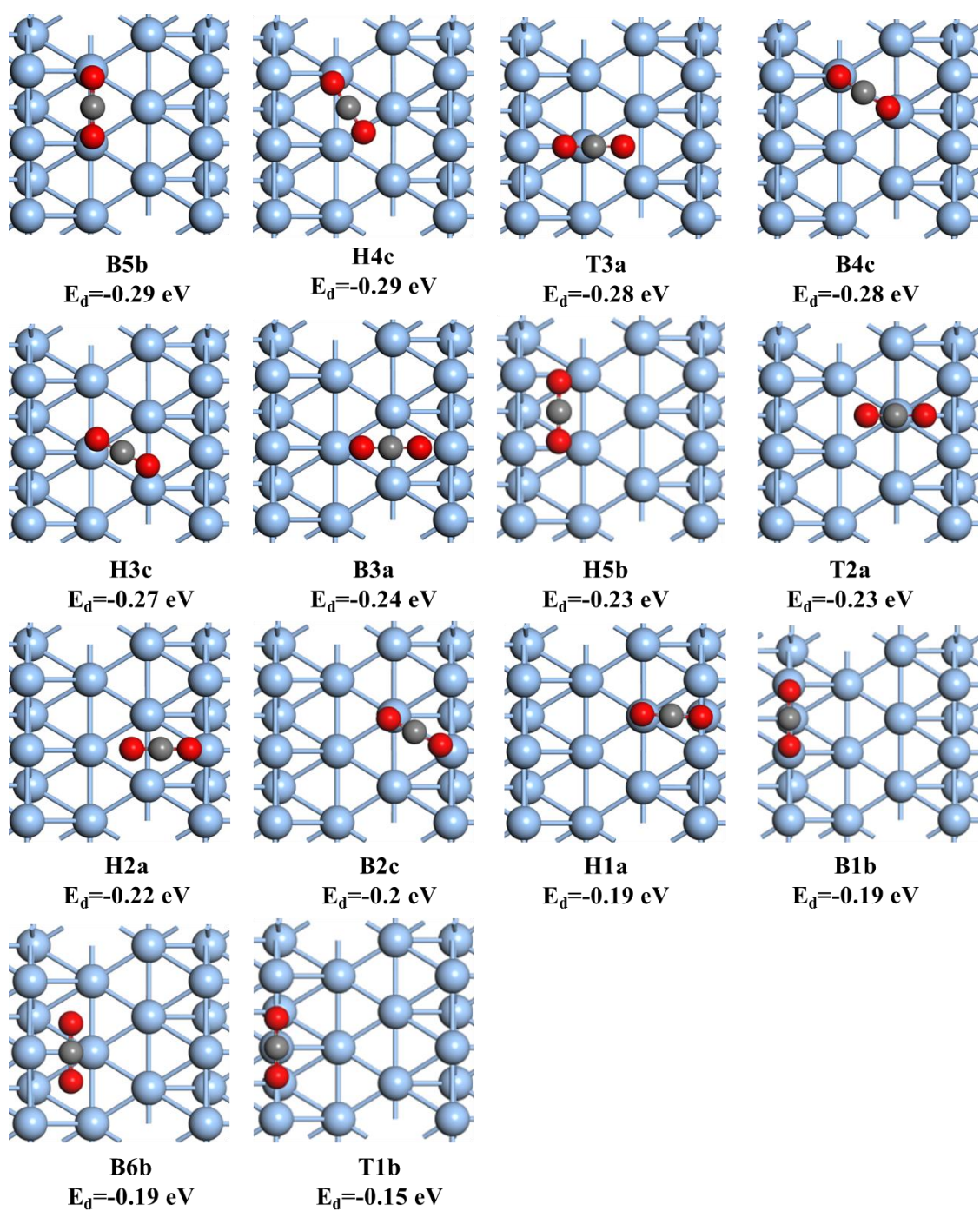

| B5b | H4c | T3a | B4c |
| E$_d$=-0.29 eV | E$_d$=-0.29 eV | E$_d$=-0.28 eV | E$_d$=-0.28 eV |

| H3c | B3a | H5b | T2a |
| E$_d$=-0.27 eV | E$_d$=-0.24 eV | E$_d$=-0.23 eV | E$_d$=-0.23 eV |

| H2a | B2c | H1a | B1b |
| E$_d$=-0.22 eV | E$_d$=-0.2 eV | E$_d$=-0.19 eV | E$_d$=-0.19 eV |

| B6b | T1b |
| E$_d$=-0.19 eV | E$_d$=-0.15 eV |

**Figure. 3**. Top views of optimized structures of the (211) surface, which are sorted by the adsorption energies. Color legend: Ag grey, C black and O red.

Based on the DFT calculations, Wang et al. [27] found that the adsorption energies are somehow dependent on the coverage of $CO_2$. Here, we selected four representative sites (*i.e.*, T1, T2, T3, and B1 of the (211) surface), and changed the coverage from 1/9 to 1/18. As shown in Table S4, adsorption energy differences are negligible, which are less than 0.02 eV. Importantly, the order of the adsorption energies remained the same. Moreover, we additionally considered structural models with the $CO_2$ having chemical interactions with the surface (distances between $CO_2$ and the (211) surface are set as 2.2 Å, and ∠OCO as 120° and 180°, see Fig. S5). The DFT calculations show that the structures and the adsorption energies are almost the same to that depicted in Fig. 3 and Fig. S4 (see Table S5).

The selected optimized structures and corresponding adsorption energies of $CO_2$ on the second commonly studies (110) surface are shown in Fig. 4, and the full list of the optimized structures are given in Fig. S6, with the corresponding adsorption energies listing in Table S6. Similar to that of the (211) surface, the adsorption energies are strongly dependent on the adsorption sites. Generally, the adsorption energies at T2, B3, H2 and B2 sites are relatively larger, and these adsorption positions are either at or close to the terrace sites, which are marked with the red dotted box in Fig. 1b. The positions lie on the step sites have relatively smaller adsorption energies (*i.e.* H1, B1 and T1). In the stable structures (*i.e.* T2a, B3a, and H2a in Fig. 4), the distances between C and the nearest Ag are 3.7 Å, 3.5 Å, 3.4 Å, respectively, and distances from O to the nearest Ag are 3.2 Å (3.3 Å), 2.9 Å, 3.0 Å (3.2 Å), respectively. The values of $d_{O-Ag}$ and $d_{C-Ag}$ of all structures are listed in Table S7. Moreover, the Bader charge analysis shows charge transfers between $CO_2$ and the (110) surface are 0.058 $e$ for T2a, 0.053 $e$ for B2a, and 0.046 $e$ for B1b. Based on these calculations, we conclude that the adsorption of $CO_2$ on the (110) surface are again the physical adsorption, which is the same to that of the (211) surface. Moreover, previous studies[36, 51] also reported the weak interactions of $CO_2$ with the (110) surface by the electron energy loss spectroscopy (EELS), low energy electron diffraction (LEED) and temperature programmed desorption (TPD).

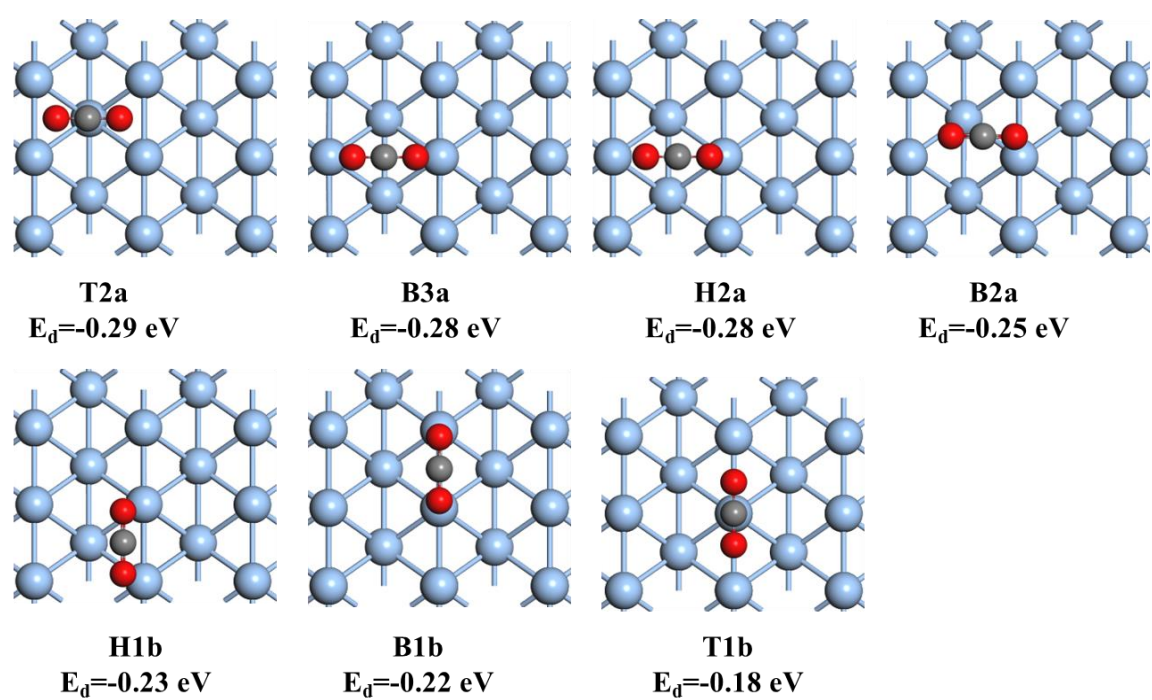

**Figure. 4.** Top view of optimized structures of the (110) surface, which are sorted by the adsorption energies. Color legend: Ag grey, C black and O red.

For the $CO_2$ on the (111) surface, we optimized thirteen adsorption structures (see Fig. S7). The most stable structures and corresponding adsorption energies are shown in Fig. 5. DFT calculations show that adsorption energies at B, H and F sites are similar (-0.26 eV), which are higher than that at the T site (-0.23 eV). The largest adsorption energy at B, H and F sites on the (111) surface is -0.26 eV, and the Bader charge analysis shows charge transfers between $CO_2$ and (111) surface is 0.055 $e$ for Fa and Ha, and 0.051 $e$ for Ta, respectively. These calculations indicate that the adsorption of

$CO_2$ on the (111) surface belongs to the physical adsorption, which is similar to the previous experimental results obtained by the low energy electron diffraction (LEED).[52] Moreover, nine structures were considered for the $CO_2$ on the (100) surface (Fig. S8), and the most stable adsorption structures and corresponding adsorption energies are shown in Fig. 5. We found that the order of adsorption energies at follow an order of H>B>T, and the most stable structure on the (100) surface is Ha, having an adsorption energy of -0.26 eV. The Bader charge analysis also confirmed the physical adsorption characteristics of the $CO_2$ on the (100) surface. For example, the corresponding distances of O and C to the nearest Ag for the Ha structure ($d_{C-Ag}$ = 3.6 Å, $d_{O1-Ag}$ = 3.3 Å, $d_{O2-Ag}$ = 3.3 Å). The Bader charge analysis shows charge transfers between $CO_2$ and the (100) surface are 0.049 $e$ for Ha, 0.043 $e$ for Ba, and 0.037 $e$ for Ta, respectively.

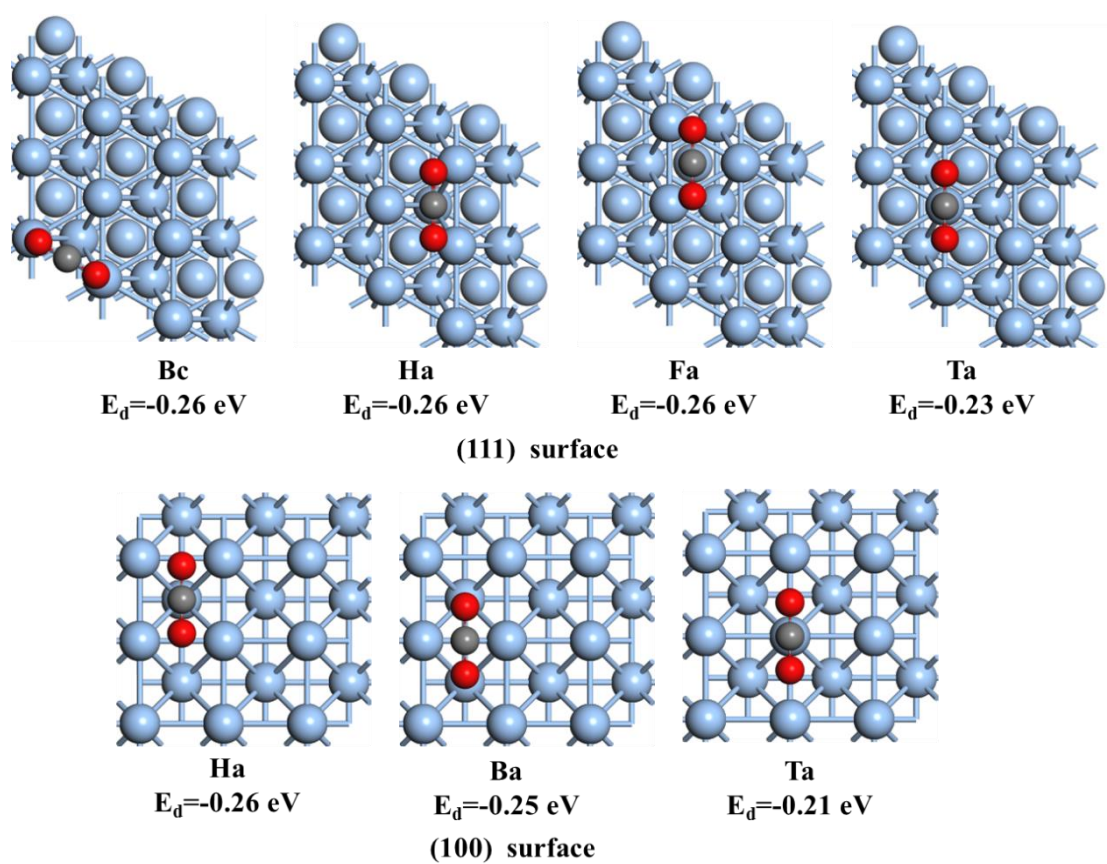

**Figure. 5**. Top view of the optimized structures of (111) and (100) surfaces, which are sorted by the adsorption energies. Color legend: Ag grey, C black and O red.

Lastly, based on the adsorption energies, we calculated the populations of $CO_2$ on different Ag surfaces by employing Boltzmann distributions (Eq. 2), and the results are given in Table 1. The results show that the populations of $CO_2$ on four surfaces are different, having the order of (211) > (110) > (111) > (100). The possibilities of locating $CO_2$ on (211) and (110) surfaces are 49.85%, and 31.7%, respectively. The summation of 81.55% indicates that these two surfaces are the most reactive facets to capture the $CO_2$. However, the other two surfaces, (111) and (100) only have possibilities of 11.91% and 6.55%, respectively to capture the $CO_2$. Moreover, the populations of $CO_2$ on different adsorption sites are also different even for the same surface. Taking the (211) surface as an example, $CO_2$ at B5 and T1 sites are 11.56 % and 0.05%, respectively. Interestingly, we found that populations of $CO_2$ at step sites (T or T1) are the lowest for all surfaces, which is contrary to the common belief where the populations at T (or T1) are the highest. [39] In contrast, the populations of $CO_2$ at the terrace sites are the highest for all studies surfaces. For example, the B5 at the (211) and T2 at the (110) surface. To further verify the rules presented in Table 1, we calculated the *d*-band center of (211), (110), (111) and (100) surfaces, which are -3.70 eV, -3.97 eV, -4.02 eV, -4.14 eV, respectively, which are similar to the reported values of -4.04 eV for the *d*-band center of (111).[53] The results of the linearly correlation between the adsorption possibilities and the positions of *d*-band center support our findings on the population of $CO_2$ on Ag surfaces.[35, 68]. Moreover, we calculated charge transfers between $CO_2$ and Ag surfaces, and we found that increasing charge transfer between $CO_2$ and Ag surfaces makes the structures more stable when $CO_2$ is adsorbed on the metal surfaces.[52]

**Table 1**. The populations of $CO_2$ at the adsorption sites of (211), (110), (111) and (100) surfaces. The corresponding optimized structures and adsorption energies are provided in Fig.3 ~Fig.5.

| (211) | | (110) | | (110) | | (111) | | (100) | |
| --- | --- | --- | --- | --- | --- | --- | --- | --- | --- |
| sites | % | sites | % | sites | % | sites | % | sites | % |
| B5 | 11.56 | T2 | 1.12 | T2 | 11.56 | B | 3.60 | H | 3.60 |
| H4 | 11.56 | H2 | 0.76 | B3 | 7.83 | H | 3.60 | B | 2.44 |
| T3 | 7.83 | B2 | 0.35 | H2 | 7.83 | F | 3.60 | T | 0.51 |
| B4 | 7.83 | H1 | 0.24 | B2 | 2.44 | T | 1.12 | -- | -- |
| H3 | 5.31 | B1 | 0.24 | H1 | 1.12 | -- | -- | -- | -- |
| B3 | 1.65 | B6 | 0.24 | B1 | 0.76 | -- | -- | -- | -- |
| H5 | 1.12 | T1 | 0.05 | T1 | 0.16 | -- | -- | -- | -- |
| Sum | | | 49.85 | | 31.70 | | 11.91 | | 6.55 |

CONCLUSIONS

As a summary, the adsorption states of $CO_2$ on (211), (110), (111), and (100) surfaces of the Ag electrode were investigated by DFT calculations. The most stable $CO_2$ adsorption sites are found for all studies as well as their adsorption structures. Adsorption populations of $CO_2$ on different surfaces are found with the order of (211) > (110) > (111) > (100). The most stable adsorption positions are not the previous reported step sites, but the terrace sites. $CO_2$ is mainly adsorbed by vdW interactions, and there is very little charge transfer (< 0.1 $e$) between $CO_2$ and Ag surfaces. The adsorption energies of $CO_2$ vary largely at different adsorption sites on the same Ag surface, which is somehow corrected with $d$-band center positions and the charge transfers between Ag surfaces and $CO_2$.

**ASSOCIATED CONTENT**

**Supporting Information**

The Supporting Information is available free of charge on the ACS Publications website.

The benchmarked calculations, adsorption energies, and optimized structures.

## AUTHOR INFORMATION

Corresponding authors:

Lei Liu, liulei@ipe.ac.cn; liulei3039@gmail.com

Suojiang Zhang, sjzhang@ipe.ac.cn

## Notes

The authors declare no competing financial interest.

## ACKNOWLEDGMENTS

This work was financially supported by the Key Program of National Natural Science Foundation of China (21978306), DNL Cooperation Fund, CAS (DNL180406) and the CAS Pioneer Hundred Talents Program (L. L).